\documentclass[preprint]{revtex4}
\usepackage{graphicx}
\newcommand{\Prob}{\mbox{Prob}}
\newcommand{\sgn}{\mbox{sgn}}

\renewcommand{\vec}[1]{\mbox{\boldmath $#1$}}
\begin{document}

\title[Transient dynamics for sequence processing neural networks]
{Transient dynamics for sequence processing neural networks}

\author{Masaki KAWAMURA}
\affiliation{ Faculty of Science, Yamaguchi University, Yoshida 1677-1,
Yamaguchi, 753-8512 Japan}
\email{kawamura@sci.yamaguchi-u.ac.jp}
\author{Masato OKADA}
\affiliation{RIKEN BSI, Wako-shi, 351-0198 Japan}

\preprint{J. Phys. A: Math. Gen., {\bfseries 35}, 253--266 (2002)}

\date{\today}

\begin{abstract}
An exact solution of the transient dynamics for a sequential
associative memory model is discussed through both the path-integral
method and the statistical neurodynamics.  Although the path-integral
method has the ability to give an exact solution of the transient
dynamics, only stationary properties have been discussed for the
sequential associative memory. We have succeeded in deriving an exact
macroscopic description of the transient dynamics by analyzing the
correlation of crosstalk noise.  Surprisingly, the order parameter
equations of this exact solution are completely equivalent to those of
the statistical neurodynamics, which is an approximation theory that
assumes crosstalk noise to obey the Gaussian distribution.  In order to
examine our theoretical findings, we numerically obtain cumulants of the
crosstalk noise.  We verify that the third- and fourth-order cumulants
are equal to zero, and that the crosstalk noise is normally distributed
even in the non-retrieval case.  We show that the results obtained by
our theory agree with those obtained by computer simulations.  We have
also found that the macroscopic unstable state completely coincides with
the separatrix.  
\end{abstract}


\maketitle

\section{Introduction}

Statistical mechanical theories have been applied to the field of neural
networks such as combinatorial optimization and associative
memories. Recently, they have also been applied to the field of
information science such as error-correcting codes, image recovery, and
CDMA
\cite{Sourlas1989,Sourlas1998,Rujan1993,MoritaTanaka1996,Nishimori1999,Iba1999,KabashimaSaad2000,T.Tanaka2000}.
For example, the maximum {\it a posteriori\/} probability (MAP) and the
maximum posterior marginals (MPM) in the framework of the Bayesian
estimation correspond to finding the ground state and the equilibrium
state of the corresponding spin system, respectively.  From this point
of view, information processing can be treated as some kind of
relaxation process of spin systems. The transient dynamics of systems
should therefore be discussed from the information theoretic point of
view.

In general, however, theoretical treatment of the dynamics is extremely
difficult compared with the equilibrium statistical mechanics of
frustrated systems into which many problems concerning the information
processing are mapped. Although a lot of works has been done to analyze
the transient dynamics \cite{Gardner1987,Amari1988,Coolen1993}, it is
well known that achieving the tractable rigorous treatment is hopeless
regarding the dynamics of frustrated systems.  In this paper, we discuss
a correlation type associative memory model having frustrated couplings,
and rigorously derive an exact macroscopic description of the transient
dynamics for this model.

There are two types of associative memory models: autoassociative memory
models and sequential associative memory models. In the case of
autoassociative memory models \cite{Hopfield1982}, their storage
capacity and phase diagram have been analyzed by equilibrium theories
\cite{Amit1987,Shiino1992}.  In the case of sequential associative
memory models, there is no equilibrium state, since the network
retrieves different patterns sequentially. As we mentioned above, since
associative memory models definitely belong to information processing
systems, it is important to analyze the information process, that is,
the retrieval process of the stored patterns. Although the path-integral
method based on the generating function
\cite{Gardner1987,Sommers1987,Gomi1995,Koyama1999} has the potential
ability to provide a rigorous solution for an associative memory model,
the theory is formal and intractable because the complexity of the
numerical calculation is exponentially large with respect to the time
step.  Accordingly, the theory can only describe the short time region
and equilibrium state.

Recently, D\"uring et al. \cite{During1998} presented the path-integral
method regarding a sequential associative memory model. However, they
did not discuss the transient properties of the retrieval process, but
only analyzed properties of the stationary state, i.e., the storage
capacity and the phase diagram \cite{During1998}.  On the other hand,
the statistical neurodynamics
\cite{Amari1988,Patrick1991a,NishimoriOzeki1993,Okada1995,Kawamura1999}
is an approximation theory capable of analyzing the long-term behavior
of the transient dynamics.  Here, the input to the spin or neuron can be
divided into two terms.  The first one is a signal term, which is a
signal to retrieve, and the second one is a crosstalk noise term, which
prevents retrieval. In the statistical neurodynamics, one assumes that
the crosstalk noise obeys the Gaussian distribution with mean 0 and a
time dependent variance, and derives macroscopic recursive equations for
the amplitude of the signal term and the variance of the crosstalk
noise.  Therefore, the basin of attraction and other dynamic properties
can be discussed through the statistical neurodynamics.  For the
autoassociative memory model, Nishimori and Ozeki tried to numerically
check the Gaussian assumption using large-scale computer simulations,
and found that the assumption holds only when the retrieval process
succeeds \cite{NishimoriOzeki1993,Ozeki1994}.

In this paper, we derive an exact solution of the transient dynamics for
the sequential associative memory model through the path-integral
method, and derive an approximated solution through the statistical
neurodynamics.  We then compare the results of the two theories, and try
to clarify their relationship.

This paper is constructed as follows. In the second section, the
sequential associative memory model that we use is defined. In the third
section, we introduce the formal macroscopic state equations obtained by
D\"uring et al. using the path-integral method. In the fourth section,
we derive an exact macroscopic description of the transient dynamics by
the path-integral method. In the fifth section, we also derive
macroscopic state equations by the statistical neurodynamics, and prove
that both the path-integral method and the statistical neurodynamics
give the same equations for the sequential associative memory model. In
the sixth section, the transient dynamics are verified by comparing
these theories with computer simulations.

\section{Sequence Processing Neural Network}

Let us consider a sequential associative memory model that consists
of $N$ spins or neurons. We consider the case of $N \to\infty$.  The
state of the spins takes $\sigma_i(t)=\pm1$ and updates the state
synchronously with the following probability:
\begin{eqnarray}
 \Prob\left[\sigma_i(t+1)\vert h_i(t)\right]
  &=& \frac12\left[1+\sigma_i(t+1)\tanh\beta h_i(t)\right] ,
  \label{eqn:dynamics} \\
 h_i(t) &=& \sum_{j=1}^NJ_{ij}\sigma_j(t)+\theta_i(t) ,
  \label{eqn:hi}
\end{eqnarray}
where $\beta$ is the inverse temperature, $\beta=1/T$. When the
temperature is absolute zero, i.e., $T=0$, the state of the spins
$\sigma_i(t+1)$ is determined by the sign of local field $h_i(t)$, that
is,
\begin{eqnarray}
 \sigma_i(t+1)=\sgn\left[h_i(t)\right] .
  \label{eqn:sgn_h}
\end{eqnarray}
The term $\theta_i(t)$ is a threshold or an external input to the
network. Synaptic connection $J_{ij}$ stores $p$ random patterns
$\vec{\xi}^{\mu}=(\xi^{\mu}_1,\cdots,\xi^{\mu}_N)^T$ so as to retrieve
the patterns as
$\vec{\xi}^1\to\vec{\xi}^2\to\cdots\vec{\xi}^p\to\vec{\xi}^1$
sequentially. For instance, it is given by
\begin{equation}
 J_{ij}=\frac1N\sum_{\mu=1}^p\xi^{\mu+1}_i\xi^{\mu}_j ,
  \label{eqn:Jij}
\end{equation}
where $\vec{\xi}^{p+1}=\vec{\xi}^1$. The number of stored patterns $p$
is given by $p=\alpha N$, where $\alpha$ is called {\itshape the loading
rate}. Each component of the patterns is assumed to be an independent
random variable that takes a value of either $+1$ or $-1$ according to
the following probability,
\begin{equation}
 \Prob\left[\xi_i^{\mu}=\pm1\right]=\frac{1}{2} .
\end{equation}
We determine the initial state $\vec{\sigma}(0)$ according to the
following probability distribution,
\begin{equation}
 \Prob[\sigma_i(0)=\pm1] = \frac{1\pm m(0) \xi^p_i}{2},
\end{equation}
and therefore the overlap between the pattern $\vec{\xi}^p$ and the
initial state $\vec{\sigma}(0)$ is $m(0)$.  The network state
$\vec{\sigma}(t)$ at time $t$ is expected to be near the pattern
$\vec{\xi}^t$, when initial overlap $m(0)$ is large and the loading rate
is small.

\section{Path-integral Method}

D\"uring et al. \cite{During1998} discussed the sequential associative
memory model by the path-integral method. In this section, we introduce
macroscopic state equations for the model with a finite temperature 
$T\geq0$, according to their paper. The detailed derivation is available
in their paper \cite{During1998}.

In order to analyze the transient dynamics, the generating function
$Z[\vec{\psi}]$ is defined as
\begin{eqnarray}
 Z[\vec{\psi}] &=& 
 \sum_{\vec{\sigma}(0),\cdots,\vec{\sigma}(t)} 
 p\left[\vec{\sigma}(0),\vec{\sigma}(1),\cdots,\vec{\sigma}(t)\right]
 e^{-i\sum_{s<t} \vec{\sigma}(s)\cdot \vec{\psi}(s)} ,
 \label{eqn:Z0} 
\end{eqnarray}
where
$\vec{\psi}=\left(\vec{\psi}(0),\cdots,\vec{\psi}(t-1)\right)$. The
state $\vec{\sigma}(s)=(\sigma_1(s),\cdots,\sigma_N(s))^T$ denotes the
state of the spins at time $s$, and the probability
$p\left[\vec{\sigma}(0),\vec{\sigma}(1),\cdots,\vec{\sigma}(t)\right]$
denotes the probability of taking the path from initial state
$\vec{\sigma}(0)$ to state $\vec{\sigma}(t)$ at time $t$ through
$\vec{\sigma}(1),\vec{\sigma}(2),\cdots,\vec{\sigma}(t-1)$. As
(\ref{eqn:Z0}) shows, the generating function takes the summation of all
$2^{N(t+1)}$ paths, which the network can take from time 0 to $t$. The
generating function $Z[\vec{\psi}]$ involves the sequence overlap
$m(s)$, which represents the direction cosine between the state
$\vec{\sigma}(s)$ and the retrieval pattern $\vec{\xi}^s$ at time $s$,
the response functions $G(s,s')$, and the correlation functions
$C(s,s')$ as follows:
\begin{eqnarray}
 m(s) &=& i\lim_{\vec{\psi}\to0}\frac{1}{N}\sum_{i=1}^N \xi_i^s
  \frac{\partial Z[\vec{\psi}]}{\partial \psi_i(s)} 
 = \frac{1}{N}\sum_{i=1}^N \xi_i^s\left<\sigma_i(s)\right> ,
 \label{eqn:def_m} \\
 G(s,s') &=& i\lim_{\vec{\psi}\to0}\frac{1}{N}\sum_{i=1}^N
  \frac{\partial^2 Z[\vec{\psi}]}{\partial\psi_i(s)\partial\theta_i(s')} 
  = \frac{1}{N}\sum_{i=1}^N\frac{\partial\left<\sigma_i(s)\right>} 
  {\partial\theta_i(s')} ,
  \\
 C(s,s') &=& -\lim_{\vec{\psi}\to0}\frac{1}{N}\sum_{i=1}^N
  \frac{\partial^2 Z[\vec{\psi}]}{\partial\psi_i(s)\partial\psi_i(s')} 
  = \frac{1}{N}\sum_{i=1}^N\left<\sigma_i(s)\sigma_i(s')\right> ,
  \label{eqn:def_C}
\end{eqnarray}
where $\left<\cdot\right>$ denotes the thermal average. Using the
assumption of self-averaging, we replace the generating function
$Z[\vec{\psi}]$ with its ensemble average $\overline{Z}[\vec{\psi}]$.

Evaluating the averaged generating function
$\overline{Z}[\vec{\psi}]$ through the saddle point method, D\"uring et
al. succeeded in obtaining the following macroscopic recursive equations
for the order parameters of (\ref{eqn:def_m})--(\ref{eqn:def_C}).
\begin{eqnarray}
 m(s) &=& \left<\xi^{s}\int \left\{d\vec{v}d\vec{w}\right\}
	   e^{i\vec{v}\cdot\vec{w}-\frac12\vec{w}\cdot\vec{R}\vec{w}}
	   \tanh \beta\left[\xi^sm(s-1)+\theta(s-1)+\sqrt{\alpha}v(s-1)\right]
	       \right>_{\xi} \label{eqn:M_w} , \nonumber \\ \\
 G(s,s') &=& \delta_{s,s'+1}\beta
  \left\{1-\left< \int\left\{d\vec{v}d\vec{w}\right\}
	    e^{i\vec{v}\cdot\vec{w}-\frac12\vec{w}\cdot\vec{R}\vec{w}}
	    \right. \right. \nonumber \\
 && \left. \left. \times
	    \tanh^2\beta\left[\xi^sm(s-1)+\theta(s-1)+\sqrt{\alpha}v(s-1)
		       \right]\right>_{\xi}\right\}  \label{eqn:G_w} , \\
 C(s,s') &=& \delta_{s,s'}+\left(1-\delta_{s,s'}\right)
  \left< \int \left\{d\vec{v}d\vec{w}\right\}
  e^{i\vec{v}\cdot\vec{w}-\frac12\vec{w}\cdot\vec{R}\vec{w}}
  \right. \nonumber \\
 && \times \tanh\beta \left[\xi^sm(s-1)+\theta(s-1)+\sqrt{\alpha}v(s-1)\right]
  \nonumber \\
 && \left. \times \tanh\beta\left[\xi^{s'}m(s'-1)+\theta(s'-1)
			     +\sqrt{\alpha}v(s'-1)\right] \right>_{\xi}
 \label{eqn:C_w} ,
\end{eqnarray}
where $\left\{d\vec{v}d\vec{w}\right\}=\prod_{s<t}
\left[\frac{dv(s)}{\sqrt{2\pi}}\frac{dw(s)}{\sqrt{2\pi}}\right]$ for
$\vec{v}=(v(0),v(1),\cdots,v(t-1))^T$,
$\vec{w}=(w(0),w(1),\cdots,w(t-1))^T$, $\left<\cdot\right>_{\xi}$
denotes the average over all $\xi$'s, and the matrix $\vec{R}$ is given
by
\begin{equation}
 \vec{R}=\sum_{n\geq0}\left[\left(\vec{G}\right)^n\vec{C}
		       \left(\vec{G}^{\dag}\right)^n\right]
		       \label{eqn:R} .
\end{equation}

\section{Dynamic treatment of the path-integral method}
\subsection{Transient dynamics}

These formal dynamical equations seem to be intractable because the
numerical complexity in directly calculating these equations becomes
exponentially large with respect to the time $s$.  Although the rigorous
solution of these equations formally has an ability to treat the
macroscopic state transition, D\"uring et al. derived stationary state
equations from these formal dynamical equations, and only analyzed the
stationary properties of the storage capacity and the phase diagram.  In
contrast, we have succeeded in obtaining a tractable description of the
macroscopic dynamic state transition from the result as shown below.

Since the matrix $\vec{R}$ is $R(s,s')=\left<v(s)v(s')\right>$, the
matrix $\vec{R}$ represents the covariance matrix of crosstalk noise
$\vec{v}$. We need the value of $R(s,s')$ for each time to analyze the
transient dynamics exactly. In consideration of this, we reconsidered
(\ref{eqn:G_w}) and succeeded in deriving a recurrence relation form of
$R(s,s')$. From (\ref{eqn:G_w}), $G(s,s')=0$ is satisfied when $s\neq
s'+1$. Therefore, the matrix $\vec{G}$ can be given by
\begin{eqnarray}
 \vec{G} &=& \left[ 
	      \begin{array}{ccccc}
	       0  & 0 & 0& \cdots & 0 \\
	       g_1 & 0 & 0 & \cdots & 0 \\
	       0 & g_2 & 0 & \cdots & 0 \\
	       \vdots & & \ddots & & \vdots \\
	       0 & \cdots & 0 & g_{t-1} & 0\\
	      \end{array}
	    \right] ,
\end{eqnarray}
with $g_{s}=G(s,s-1)$. From this, we can easily find that
$G^n(s,s')=\delta_{s,s'+n}\prod_{\tau=0}^{n-1}g_{s-\tau}, (n\ge1)$ and
\begin{eqnarray}
 \left[\left(\vec{G}\right)^n \vec{C} 
  \left(\vec{G}^{\dag}\right)^n\right](s,s') &=& \left\{
  \begin{array}{ll}
   C(s,s') & , n=0 \\
   \displaystyle{C(s-n,s'-n)\prod_{\tau=0}^{n-1}g_{s-\tau}
    \prod_{\tau'=0}^{n-1}g_{s'-\tau'}}
    & , 1\le n\le s\\
   0 & , n>s \\
  \end{array}
       \right. .
\end{eqnarray}
From (\ref{eqn:R}), $R(s,s')$ can be reduced to
\begin{eqnarray}
 \lefteqn{R(s,s') = C(s,s')+ \sum_{n\ge1}^{s} C(s-n,s'-n) 
  \prod_{\tau=0}^{n-1}G(s-\tau,s-\tau-1) 
  \prod_{\tau'=0}^{n-1}G(s'-\tau',s'-\tau'-1)} \nonumber \\
 && \\
 &=& C(s,s')+C(s-1,s'-1)G(s,s-1)G(s',s'-1) \nonumber \\
 && +\sum_{n\ge1}^{s-1} C(s-n-1,s'-n-1) 
  \prod_{\tau=0}^{n}G(s-\tau,s-\tau-1)
  \prod_{\tau'=0}^{n}G(s'-\tau',s'-\tau'-1) \\
 &=& C(s,s')+G(s,s-1)G(s',s'-1)
  \left\{C(s-1,s'-1) +\sum_{n\ge1}^{s-1} C(s-n-1,s'-n-1) \right. \nonumber \\
 && \left. \times \prod_{\tau=0}^{n-1}G(s-\tau-1,s-\tau-2)
  \prod_{\tau'=0}^{n-1}G(s'-\tau'-1,s'-\tau'-2) \right\}
\end{eqnarray}
We can therefore derive a recurrence relation form of $R(s,s')$:
\begin{equation}
 R(s,s') = C(s,s')+G(s,s-1)G(s',s'-1)R(s-1,s'-1) .
  \label{eqn:R_C_G2R}
\end{equation}
Using this recurrence relation, we can evaluate the value of $R(s,s')$
for each time.

Next, since the terms $\tanh(\cdot)$ of $m(s)$ and $G(s,s-1)$ include
only the variable $v(s-1)$, these multiple integral equations can be
reduced to the following single integral equations by using
(\ref{eqn:integral}) in the Appendix\ref{ap:int_MG}.
\begin{eqnarray}
 m(s) &=& \left<\xi^s\int Dz\tanh\beta
	   \left[\xi^{s}m(s-1)+\theta(s-1)+z\sqrt{\alpha R(s-1,s-1)}\right]
	   \right>_{\xi} \label{eqn:ms} \\
 G(s,s-1) &=& \beta
  \left\{1-\left<\int Dz\tanh^2\beta
	    \left[\xi^{s}m(s-1)+\theta(s-1)+z\sqrt{\alpha R(s-1,s-1)}
	   \right]\right>_{\xi}\right\} , \nonumber \\
	    \label{eqn:G}
\end{eqnarray}
with the familiar abbreviation
$Dz=\frac{dz}{\sqrt{2\pi}}e^{-\frac{1}{2}z^2}$.  Since $C(s,s')$
includes $v(s-1)$ and $v(s'-1)$, $C(s,s')$ can be reduced to a double
integral equation from (\ref{eqn:integral}),
\begin{eqnarray}
 C(s,s) &=& 1 \\
 C(s,0) &=& \left< \xi^p m(0)
  \int Dz\tanh\beta
   \left[\xi^sm(s-1)+\theta(s-1)+z\sqrt{\alpha R(s-1,s-1)}\right]
  \right>_{\xi} \\ 
 &=& 0 \\
 C(s,s') &=& 
  \left<\int
   \frac{d\vec{z}e^{-\frac12\vec{z}\cdot\vec{R}_{11}^{-1}\vec{z}}}
   {2\pi\vert\vec{R}_{11}\vert^{\frac12}}
   \tanh\beta \left[\xi^sm(s-1)+\theta(s-1)+\sqrt{\alpha}z(s-1)\right]
 \right. \nonumber \\
 && \times \left.
 \tanh\beta\left[\xi^{s'}m(s'-1)+\theta(s'-1)+\sqrt{\alpha}z(s'-1)\right]
 \right>_{\xi} \label{eqn:Css}
\end{eqnarray}
where the matrix $\vec{R}_{11}$ is a $2\times2$ matrix consisting of
the elements of $\vec{R}$ at time $s-1$ and time $s'-1$, and
$\vec{z}=[z(s-1), z(s'-1)]^T$. We therefore have the macroscopic
state equations (\ref{eqn:R_C_G2R})--(\ref{eqn:Css}) for each time, and
can analyze the transient dynamics exactly.

\subsection{Stationary state equations}

Let us derive stationary state equations from our macroscopic state
equations.  We assume $m(t)\to m$ and $R(t,t')\to r$ when
$t,t'\to\infty$. In this case, we can get
\begin{eqnarray}
 m &=& \left<\xi\int Dz\tanh\beta
        \left[\xi m+\theta+z\sqrt{\alpha r}\right] \right>_{\xi} ,
  \label{eqn:stat_m} \\
 q &=& \left<\int Dz\tanh^2\beta
        \left[\xi m+\theta+z\sqrt{\alpha r}\right] \right>_{\xi} ,
\end{eqnarray}
and also $G(t,t-1)\to\beta\left(1-q\right)$. We can, therefore, obtain
\begin{eqnarray}
 r &=& \frac1{1-\beta^2\left(1-q\right)^2} .
  \label{eqn:stat_r}
\end{eqnarray}
These equations (\ref{eqn:stat_m})--(\ref{eqn:stat_r}) are coincident
with those obtained by D\"uring el al.  Using
(\ref{eqn:stat_m})--(\ref{eqn:stat_r}), the stationary state can be
evaluated.

\section{Statistical Neurodynamics}

\subsection{Finite temperature case}

Macroscopic state equations have also been derived for a sequential
associative memory model with absolute zero temperature $T=0$ by the
statistical neurodynamics \cite{Amari1988b}. In this paper, we derive
the macroscopic state equations for the present model with finite temperature
$T\geq0$, by the statistical neurodynamics \cite{NishimoriOzeki1993}.
In the statistical neurodynamics, the input is divided into a signal
term and a crosstalk noise term. From (\ref{eqn:hi}) and
(\ref{eqn:Jij}), we obtain
\begin{eqnarray}
 h_i(s) &=& \xi^{s+1}_im^{s}(s) +z_i(s) +\theta_i(s) ,
  \label{eqn:his} \\
 z_i(s) &=& \sum_{\mu\neq s}^p \xi^{\mu+1}_im^{\mu}(s) , \label{eq.z} \\
 m^{\mu}(s) &=& \frac1N\sum_i\xi^{\mu}_i\sigma_i(s).
\end{eqnarray}
The overlaps $ m^{\mu}(s)$ are defined for each pattern
$\vec{\xi}^{\mu}$. We assume that the crosstalk noise $z_i(s)$ is
normally distributed with mean $0$ and variance $\rho^2(s,s)$. Then, the
sequence overlap $m(s)=m^s(s)$ for condensed patterns becomes
\begin{eqnarray}
 m(s) &=& \left<\xi^s\int Dz\tanh\beta
           \left[\xi^s m(s-1)+\theta(s-1)+z\rho(s-1,s-1)\right]\right>_{\xi} .
 \label{eqn:SND_m} 
\end{eqnarray}

Next, let us evaluate the overlap $m^{\mu}(s),\mu \neq s$ between the
network state $\vec{\sigma}(s)$ and the uncondensed pattern
$\vec{\xi}^\mu$ in order to calculate the covariance matrix
$\rho^2(s,s')=E\left[z_i(s)z_i(s')\right]$ of the crosstalk noise of
(\ref{eq.z}). We assume that the pattern $\xi^{\mu}_i$ and the state
$\sigma_i(s)$ are independent regarding $i$.  The state $\sigma_i(s)$ is
correlated with the uncondensed pattern $\xi^\mu_i$, because the local
field $h_i(s-1)$ includes the pattern $\vec{\xi}^\mu$.  From
(\ref{eqn:his}), since the overlap $m^{\mu}(s)$ is $O(1/\sqrt{N})$, the
term $\tanh\left[\beta h_i(s-1)\right]$, which determines the state
$\sigma_i(s)$ stochastically, is expanded as follows:
\begin{eqnarray}
 \tanh\left[\beta h_i(s-1)\right] &=& 
  \tanh\left[\beta h_i^{(\mu)}(s-1)\right]+\beta\xi^{\mu}_im^{\mu-1}(s-1)
  \mbox{ sech}^2\left[\beta h_i^{(\mu)}(s-1)\right] , \\
 h_i^{(\mu)}(s-1) &=& \xi^{s}_im^{s-1}(s-1) 
  +\frac{1}{N}\!\! \sum_{\nu\neq s,\mu-1}^p
  \xi^{\nu+1}_im^{\nu}(s-1) +\theta_i(s-1) .
\end{eqnarray}
Here, we use $\left(\tanh x\right)'=1-\tanh^2x=\mbox{sech}^2x$. 
Then, we get
\begin{eqnarray}
 \Prob\left[\sigma_i(s)\vert h_i(s-1)\right]
  &=& \Prob\left[\sigma_i(s)\vert h_i^{(\mu)}(s-1)\right] 
  \nonumber \\
 &+& \frac12\left[1+\beta\xi^{\mu}_i\sigma_i(s)m^{\mu-1}(s-1)
          \mbox{ sech}^2\beta h_i^{(\mu)}(s-1) \right] .
\end{eqnarray}
We can derive 
\begin{eqnarray}
 m^{\mu}(s) &=& 
  \frac1N\sum_i\xi^{\mu}_i\sigma_i^{(\mu)}(s)+U(s)m^{\mu-1}(s-1).
  \label{eqn:m_m}
\end{eqnarray}
Note that $\sigma_i^{(\mu)}(s)$ is independent of $\xi^{\mu}_i$, and
$U(s)$ is given by
\begin{eqnarray}
 U(s) &=& \beta\left<\int Dz\mbox{ sech}^2\beta
                \left[\xi^{s}m(s-1)+\theta(s-1) 
                 +z\rho(s-1)\right] \right>_{\xi}  \\
 &=& \beta\left\{1-\left<\int Dz\tanh^2\beta
                \left[\xi^{s}m(s-1)+\theta(s-1) 
                 +z\rho(s-1)\right] \right>_{\xi}\right\} .
		\label{eqn:U}
\end{eqnarray}
Therefore, the covariance of the crosstalk noise is given by
\begin{eqnarray}
 \rho^2(s,s') &=& 
  \alpha E\left[\frac{1}{N}\sum_{i=1}^N
           \sigma_i^{(\mu)}(s)\sigma_i^{(\mu)}(s')\right] 
  +U(s)U(s') E\left[\sum_{\mu\neq s,s'}^p 
	       m^{\mu-1}(s-1)m^{\mu-1}(s'-1)\right] \nonumber \\
 && +U(s') E\left[\frac1N\sum_{\mu\neq s,s'}^p\sum_i
	    \xi^{\mu}_i\sigma_i^{(\mu)}(s)m^{\mu-1}(s'-1)\right] 
  \nonumber \\
 && +U(s) E\left[\frac1N\sum_{\mu\neq s,s'}^p\sum_i
	   \xi^{\mu}_i\sigma_i^{(\mu)}(s')m^{\mu-1}(s-1)\right] 
  \label{eqn:rho_expand} .
\end{eqnarray}
From (\ref{eqn:m_m}), the third term on the RHS of
(\ref{eqn:rho_expand}) becomes
\begin{eqnarray}
 \lefteqn{U(s')E\left[\frac1N\sum_{\mu\neq s,s'}^p\sum_i
   \xi^{\mu}_i\sigma_i^{(\mu)}(s) m^{\mu-1}(s'-1)\right] }
 \nonumber \\
 &=& U(s')E\left[\frac1{N^2}\sum_{\mu\neq s,s'}^p\sum_i
       \xi^{\mu}_i\xi^{\mu-1}_i
       \sigma_i^{(\mu)}(s) \sigma_i^{(\mu-1)}(s'-1) \right] 
 \nonumber \\
 && +U(s')U(s'-1)E\left[\frac1{N}\sum_{\mu\neq s,s'}^p\sum_i
	     \xi^{\mu}_i\sigma_i^{(\mu)}(s) m^{\mu-2}(s'-2) \right]. 
 \label{eqn:Esigma}
\end{eqnarray}
According to the literature \cite{KitanoAoyagi1998}, (\ref{eqn:Esigma})
becomes zero as follows. Since $\xi^{\mu}_i$ and $\xi^{\mu-1}_i$ are
independent of $\sigma_i^{(\mu)}(s)$ and $\sigma_i^{(\mu-1)}(s'-1)$,
respectively, the first term on the RHS of (\ref{eqn:Esigma}) becomes
$E\left[\xi^{\mu}_i\xi^{\mu-1}_i\sigma_i^{(\mu)}(s)
\sigma_i^{(\mu-1)}(s'-1)\right]=0$.
%%%%%
Using (\ref{eqn:m_m}) to (\ref{eqn:Esigma}) up to the initial time
iteratively, (\ref{eqn:Esigma}) becomes
\begin{eqnarray}
 U(s')E\left[\frac1N\sum_{\mu\neq s,s'}^p\sum_i
   \xi^{\mu}_i\sigma_i^{(\mu)}(s) m^{\mu-1}(s'-1)\right] 
 &=& \prod_{\tau=1}^{s'}U(\tau)
 E\left[\frac1{N}\sum_{\mu\neq s,s'}^p\sum_i
   \xi^{\mu}_i\sigma_i^{(\mu)}(s) m^{\mu-s'}(0) \right]. 
 \nonumber \\
\end{eqnarray}
Because of the sequential associative memory with $p=\alpha N$ period,
the correlations for $s'=p, 2p, 3p\cdots$ remain. These can however be
neglected in the limit $N\to\infty$. Therefore, 
%%%%%
the third term on the RHS of (\ref{eqn:rho_expand})
becomes zero. Similarly, the forth term also becomes zero.  Finally, 
the covariance can be given by
\begin{eqnarray}
 \rho^2(s,s') &=& \alpha C(s,s')+U(s)U(s') \rho^2(s-1,s'-1) .
  \label{eqn:SND_z}
\end{eqnarray}

\subsection{Path-integral method and statistical neurodynamics}

Let us compare the results obtained by the path-integral method and the
results obtained by the statistical neurodynamics. From
(\ref{eqn:R_C_G2R}) and (\ref{eqn:SND_z}), let $\rho^2(s,s')$ correspond
to $\alpha R(s,s')$. In this case, we obtain $U(s)=G(s,s-1)$ from
(\ref{eqn:G}) and (\ref{eqn:U}), and then $\rho^2(s,s')=\alpha R(s,s')$
from (\ref{eqn:R_C_G2R}) and (\ref{eqn:SND_z}).  Moreover, the overlap
from (\ref{eqn:ms}) is equal to one from (\ref{eqn:SND_m}).

As stated above, both theories give the same macroscopic state equations
for the sequential associative memory model. This finding means that the
crosstalk noise in the present model is normally distributed even if the
network fails in retrieval, and also that the macroscopic state
equations obtained by the statistical neurodynamics can give the exact
solution.

\section{Retrieval Process}

In this section, let us discuss the transient dynamics of the sequential
associative memory model. Since macroscopic state equations obtained
by the statistical neurodynamics are equivalent to those by the
path-integral method, we use the notations of $m(t),U(t),$ and
$\rho^2(t,t')$. Let $\rho^2(t,t)$ be $\rho^2(t,t)=\alpha r(t)$ when
$t'=t$.  We analyze the case of $\theta(t)=0$ 
and derive the macroscopic state equations,
\begin{eqnarray}
 m(t+1) &=& \left\langle\xi^{t+1}
	     \int Dz\tanh\beta\left[\xi^{t+1}m(t)+z\sqrt{\alpha r(t)}\right]
\right\rangle_{\xi} \label{eqn:m_fin} \\
 U(t+1) &=& \beta\left\{1-\left\langle\int Dz\tanh^2\beta
			    \left[\xi^{t+1}m(t)+z\sqrt{\alpha r(t)}\right]
		 \right\rangle_{\xi}\right\}, \\
 r(t+1)  &=&  1+U^2(t+1)r(t) .
  \label{eqn:r_fin}
\end{eqnarray}

Figure~1 shows the time evolution of the overlap $m(t)$. The solid lines
represent results obtained by the macroscopic state equations
(\ref{eqn:m_fin})--(\ref{eqn:r_fin}), and the broken lines represent
results obtained by computer simulations with $N=100000$, where the
loading rate is $\alpha=0.20,0.26$ and the inverse temperature is
$\beta=5$ ($T=0.2$).  The abscissa denotes the time $t$, and the
ordinate denotes the overlap $m(t)$.  In this case, the storage capacity
is $\alpha_c=0.246$. As shown in Fig.~1, our theory can describe the
transient dynamics quantitatively even if the network fails in retrieval
or the loading rate $\alpha$ exceeds the storage capacity $\alpha_c$.

In the autoassociative memory models, the crosstalk noise is not
normally distributed when the network fails in retrieval
\cite{NishimoriOzeki1993,Ozeki1994}. In the present sequential
associative memory model, however, the macroscopic state equations
obtained by the path-integral method are represented in the form of a
single or double Gaussian integral. Therefore, the first, second, third,
and fourth cumulants $C_1(t),C_2(t),C_3(t),C_4(t)$ are evaluated in
order to verify the distribution of the crosstalk noise. When the
crosstalk noise is normally distributed, the third and fourth cumulants
are zero. The cumulants are defined by
\begin{eqnarray}
 C_1(t) &=& \overline{z}(t) , \\ 
 C_2(t) &=& \overline{z^2}(t)-\overline{z}^2(t) , \\
 C_3(t) &=& \overline{z^3}(t)-3\overline{z}(t)\overline{z^2}(t)
  +2\overline{z}^3(t) , \\
 C_4(t) &=& \overline{z^4}(t)-3\left(\overline{z^2}(t)\right)^2
  -4\overline{z}(t)\overline{z^3}(t)
  +12\overline{z}^2(t)\overline{z^2}(t)-6\overline{z}^4(t) ,
\end{eqnarray}
where $\overline{z^n}(t)$ denotes the $n$th-order moment for the crosstalk
noise $z_i(t)$ and is defined as
\begin{equation}
 \overline{z^n}(t)=\frac{1}{N}\sum_{i=1}^N \left\{z_i(t)\right\}^n .
\end{equation}
The cumulants $C_1(t)$ and $C_2(t)$ represent the average and variance
of $z(t)$, respectively.

Figure~2 shows the time evolution of cumulants and overlap $m(t)$, where
the loading rate is $\alpha=0.20,0.26$ and the inverse temperature is
$\beta=5$ ($T=0.2$). The initial overlap is $m(0)=0.2$, where the
network fails in retrieval. The solid lines represent the overlap $m(t)$
and the variance of the crosstalk noise $r(t)$ obtained by our
theory. These lines agree with the overlap and the cumulant $C_2(t)$
obtained by the computer simulations. Since the third and fourth
cumulants are zero, we can find that the crosstalk noise is normally
distributed. In the sequential associative memory model, the third and
fourth cumulants hold $C_3(t)=C_4(t)=0$, as obtained by the
path-integral method, even if the network fails in retrieval. The
Figure~2 shows that the Gaussian assumption in the statistical
neurodynamics folds for the present model.

Figure~3 shows the transition of the overlap $m(t)$ and the variance of
the crosstalk noise $\alpha r(t)$. The solid lines represent results
obtained by our theory and the broken lines represent results obtained
by computer simulations with $N=100000$. The loading rate is
$\alpha=0.20$ and the inverse temperature is $\beta=5$ ($T=0.2$). As
shown in Fig.~3, there are two stable stationary states; one is
$m\approx1$, which represents a retrieval state, and the other is $m=0$,
which represents a non-retrieval state. We can therefore define the
critical overlap, which determines whether patterns will be
retrieved. When the initial overlap is larger than the critical overlap
$m_c$, stored patterns can be retrieved. Namely, as shown in Fig.~3,
there is a separatrix on the locus starting from the initial overlap
$m_c$, and the network passes through the separatrix at time $t=1$. In
general, an unstable stationary state obtained by macroscopic state
equations does not agree with a separatrix obtained by microscopic
dynamics, e.g., (\ref{eqn:dynamics}).  As shown by the cross mark in
Fig.~3, however, we can find that the unstable stationary state agrees
with the separatrix for the present model.

Figure~4 shows a basin of attraction where the inverse temperature is
$\beta=5$. The abscissa denotes the loading rate $\alpha$, and the
ordinate denotes the overlap. The solid line represents a result
obtained by theory. The lower line represents the critical overlap
$m_c$, and the upper line represents the stationary overlap $m_{\infty}$
at time $t\to\infty$, where the network starts from the initial overlap
$m(0)=1$. The intersection between the critical overlap $m_c$ and the
stationary overlap $m_{\infty}$ gives the storage capacity
$\alpha_c$. The broken lines represent the overlap $m$, which is the
unstable stationary state given by equations
(\ref{eqn:stat_m})--(\ref{eqn:stat_r}). The error bars represent the
median, and the first and third quartiles over 11 trials of the overlaps
$m_c$ and $m_{\infty}$, which are obtained by computer simulations with
$N=10000$.  As shown in this figure, the basin of attraction obtained by
the theory quantitatively coincides with that of the computer
simulations.

\section{Conclusions}

We rigorously derived dynamic macroscopic state equations for the
sequential associative memory model through the path-integral
method. This suggested that the crosstalk noise is normally distributed
for the sequential associative memory model. We also examined the
cumulants of the crosstalk noise by computer simulations, and found the
third and fourth cumulants to be zero. This finding strongly supported
the normally distributed crosstalk noise. Using this exact solution, we
obtained dynamical properties such as the critical overlap $m_c$, the
temporal behavior of the overlap, and so on. These theoretically obtained
results agreed with those of computer simulations. We also found
that the separatrix coincides with the unstable fixed point of the
macroscopic stationary state equations.

Historically, dynamic macroscopic state equations have been derived by
the statistical neurodynamics for a sequential associative memory model
with zero temperature $T=0$ assuming the crosstalk noise to obey the
Gaussian distribution \cite{Amari1988b}. We extended this theory to the
finite temperature case, and compared dynamic macroscopic state
equations by the statistical neurodynamics with those by the
path-integral method. As a result, we found both to be completely
equivalent for the sequential associative memory model. This means that
the statistical neurodynamics is exact for the sequential associative
memory model.

%%%%%
For the sequential associative memory model, we could drive the
recurrence relation of $R(s,s')$ and the dynamic macroscopic state
equations. This is the reason why no effective self-interaction, which
is caused by the time correlation of states, remains. In contrast, since
only one pattern is retrieved for an autoassociative memory model, the
effect of the intrinsic time correlation can not be avoided and the
distribution of the crosstalk noise becomes non-Gaussian. It is known
that this treatment of the autoassociative memory model is
intractable \cite{Gardner1987,Coolen0006011}.
%%%%%

%%%%%%%%%%%%% Appendix 
\appendix
\section{Multiple Gaussian Integral}
\label{ap:int_MG}

In order to calculate equations (\ref{eqn:M_w}) to (\ref{eqn:C_w}), 
it is sufficient to treat the following form of a multiple Gaussian 
integral, 
\begin{equation} 
 \int\left\{d\vec{v}d\vec{w}\right\}  
  e^{i\vec{v}\cdot\vec{w}-\frac{1}{2}\vec{w}\cdot\vec{R}\vec{w}} 
  f\left(\vec{v}_1\right), 
  \label{eq.target_integral} 
\end{equation} 
where we put $\vec{v}=(\vec{v}_1,\vec{v}_2)^T$. We define $\vec{S}$ as
the inverse matrix of $\vec{R}$. Corresponding to the representation of
$\vec{v}=(\vec{v}_1,\vec{v}_2)^T$, the matrix $\vec{R}$ and its
inversion $\vec{S}$ are represented as,
\begin{equation} 
 \vec{R}=\left( 
     \begin{array}{ll} 
      \vec{R}_{11} & \vec{R}_{12} \\ 
      \vec{R}_{21} & \vec{R}_{22} \\ 
     \end{array}\right), \quad \quad  
 \vec{S}=\left( 
     \begin{array}{ll} 
      \vec{S}_{11} & \vec{S}_{12} \\ 
      \vec{S}_{21} & \vec{S}_{22} \\ 
     \end{array}\right). 
 \label{eqn:defS}
\end{equation} 
For example, $\vec{v}_1$ and $\vec{R}_{11}$ are $v(s-1)$ and
$R(s-1,s-1)$ with equations (\ref{eqn:M_w}) and (\ref{eqn:G_w}),
respectively, while $\vec{v}_1=(v(s-1), v(s'-1))^T$ and
\begin{equation} 
 \vec{R}_{11}=\left( 
     \begin{array}{ll} 
      R(s-1,s-1)  & R(s-1,s'-1)  \\ 
      R(s'-1,s-1) & R(s'-1,s'-1) \\ 
     \end{array}\right),  
\end{equation} 
with equation (\ref{eqn:C_w}). According to the definition,
$\vec{S}\vec{R}=\vec{I}$. We obtain
\begin{eqnarray}
 \vec{S}_{11}\vec{R}_{11}+\vec{S}_{12}\vec{R}_{21} &=& \vec{I} 
  \label{eqn:SR1} \\
 \vec{S}_{21}\vec{R}_{11}+\vec{S}_{22}\vec{R}_{21} &=& \vec{0} ,
  \label{eqn:SR0}
\end{eqnarray}
where $\vec{I}$ and $\vec{0}$ are the identity and zero matrix,
respectively. Solving (\ref{eqn:SR1}) and (\ref{eqn:SR0}) with respect
to $\vec{R}_{11}$, we obtain
$\vec{R}_{11}=\left(\vec{S}_{11}-\vec{S}_{12}\vec{S}_{22}^{-1}\vec{S}_{21}\right)^{-1}$.

Accordingly, the integral of equation (\ref{eq.target_integral}) 
is obtained as, 
\begin{eqnarray} 
 \lefteqn{\int\left\{d\vec{v}d\vec{w}\right\}  
  e^{i\vec{v}\cdot\vec{w}-\frac{1}{2}\vec{w}\cdot\vec{R}\vec{w}} 
  f\left(\vec{v}_1\right) } \\ 
  &=& \int\left\{d\vec{v}d\vec{w}\right\}  
  e^{-\frac{1}{2}\left(\vec{w}-i\vec{R}^{-1}\vec{v}\right)^{\dag}\vec{R} 
  \left(\vec{w}-i\vec{R}^{-1}\vec{v}\right) 
  -\frac{1}{2}\vec{v}\cdot\vec{R}^{-1}\vec{v}} f\left(\vec{v}_1\right) \\ 
 &=& \frac{1}{\vert\vec{R}\vert^{\frac12}} \int\left\{d\vec{v}\right\} 
  e^{-\frac{1}{2}\vec{v}\cdot\vec{R}^{-1}\vec{v}} f\left(\vec{v}_1\right) \\ 
 &=& \frac{1}{\vert\vec{R}\vert^{\frac12}} \int\left\{d\vec{v}\right\} 
 e^{-\frac{1}{2}\left(\vec{v}_1\cdot \vec{S}_{11}\vec{v}_1 
                  +\vec{v}_1\cdot \vec{S}_{12}\vec{v}_2 
                  +\vec{v}_2\cdot \vec{S}_{21}\vec{v}_1 
                  +\vec{v}_2\cdot \vec{S}_{22}\vec{v}_2\right)} 
  f\left(\vec{v}_1\right) \\ 
 &=& \frac{1}{\vert\vec{R}\vert^{\frac12}} \int\left\{d\vec{v}\right\} 
  e^{-\frac{1}{2} 
  \left(\vec{v}_2+\vec{S}_{22}^{-1}\vec{S}_{12}^{\dag}\vec{v}_1\right)^{\dag} 
  \vec{S}_{22}\left(\vec{v}_2+\vec{S}_{22}^{-1}\vec{S}_{21}\vec{v}_1\right) 
  -\frac{1}{2}\vec{v}_1^{\dag}
  \left(\vec{S}_{11}-\vec{S}_{12}\vec{S}_{22}^{-1}\vec{S}_{21}\right) 
  \vec{v}_1} 
  f\left(\vec{v}_1\right) \nonumber \\ 
 && \\ 
 &=& \frac{1}{\vert\vec{R}\vert^{\frac12}\vert \vec{S}_{22}\vert^{\frac12}}  
  \int\left\{d\vec{v}_1\right\} 
  e^{-\frac{1}{2}\vec{v}_1\cdot \vec{R}_{11}^{-1} \vec{v}_1} 
  f\left(\vec{v}_1\right) ,
  \label{eqn:MultInt} 
\end{eqnarray} 
with $\left\{d\vec{v}\right\}=\prod_{s<t}
\left[\frac{dv(s)}{\sqrt{2\pi}}\right]$ for
$\vec{v}=(v(0),v(1),\cdots,v(t-1))^T$, and
$\left\{d\vec{v}_1\right\}=\frac{dv(s-1)}{\sqrt{2\pi}}$ or
$\left\{d\vec{v}_1\right\}=\frac{dv(s-1)}{\sqrt{2\pi}}
\frac{dv(s'-1)}{\sqrt{2\pi}}$. 
Substituting the following identities, 
\begin{eqnarray} 
 \vert\vec{S}\vert &=&  
  \left|\begin{array}{cc}
   \vec{S}_{11}-\vec{S}_{12}\vec{S}_{22}^{-1}\vec{S}_{21} & \vec{0} \\
	 \vec{S}_{21} & \vec{S}_{22} 
	\end{array}\right| \\
  &=& \vert \vec{S}_{11}-\vec{S}_{12}\vec{S}_{22}^{-1}\vec{S}_{21}\vert 
  \vert \vec{S}_{22}\vert  \\
 \vert\vec{S}\vert \vert\vec{R}\vert &=& 1 , 
\end{eqnarray} 
into equation (\ref{eqn:MultInt}), 
we get  
\begin{eqnarray} 
 \lefteqn{\frac{1}{\vert\vec{R}\vert^{\frac12}
  \vert \vec{S}_{22}\vert^{\frac12}}  
  \int\left\{d\vec{v}_1\right\} 
  e^{-\frac{1}{2}\vec{v}_1\cdot \vec{R}_{11}^{-1} \vec{v}_1} 
  f\left(\vec{v}_1\right) } \\ 
 &=& \vert \vec{S}_{11}-\vec{S}_{12}\vec{S}_{22}^{-1}\vec{S}_{21}\vert^{\frac12} 
  \int\left\{d\vec{v}_1\right\} 
  e^{-\frac{1}{2}\vec{v}_1\cdot \vec{R}_{11}^{-1} \vec{v}_1} 
  f\left(\vec{v}_1\right) \\ 
 &=& \frac{1}{\vert \vec{R}_{11}\vert^{\frac12}} 
  \int\left\{d\vec{v}_1\right\} 
  e^{-\frac{1}{2}\vec{v}_1\cdot \vec{R}_{11}^{-1} \vec{v}_1} 
  f\left(\vec{v}_1\right) . 
\end{eqnarray} 
We can therefore obtain a simple form of the multiple integral, 
\begin{eqnarray} 
 \int\left\{d\vec{v}d\vec{w}\right\}  
  e^{i\vec{v}\cdot\vec{w}-\frac{1}{2}\vec{w}\cdot \vec{R}\vec{w}} 
  f\left(\vec{v}_1\right)  
  &=& \frac{1}{\vert \vec{R}_{11}\vert^{\frac12}} 
  \int\left\{d\vec{v}_1\right\} 
  e^{-\frac{1}{2}\vec{v}_1\cdot \vec{R}_{11}^{-1} \vec{v}_1} 
  f\left(\vec{v}_1\right) . 
  \label{eqn:integral} 
\end{eqnarray}

%%% bibliography

%%%%%%%%%%%%%%%%%%%%
%%% Figures

%\newpage

\begin{figure*}
 \begin{center}
  \includegraphics[width=130mm]{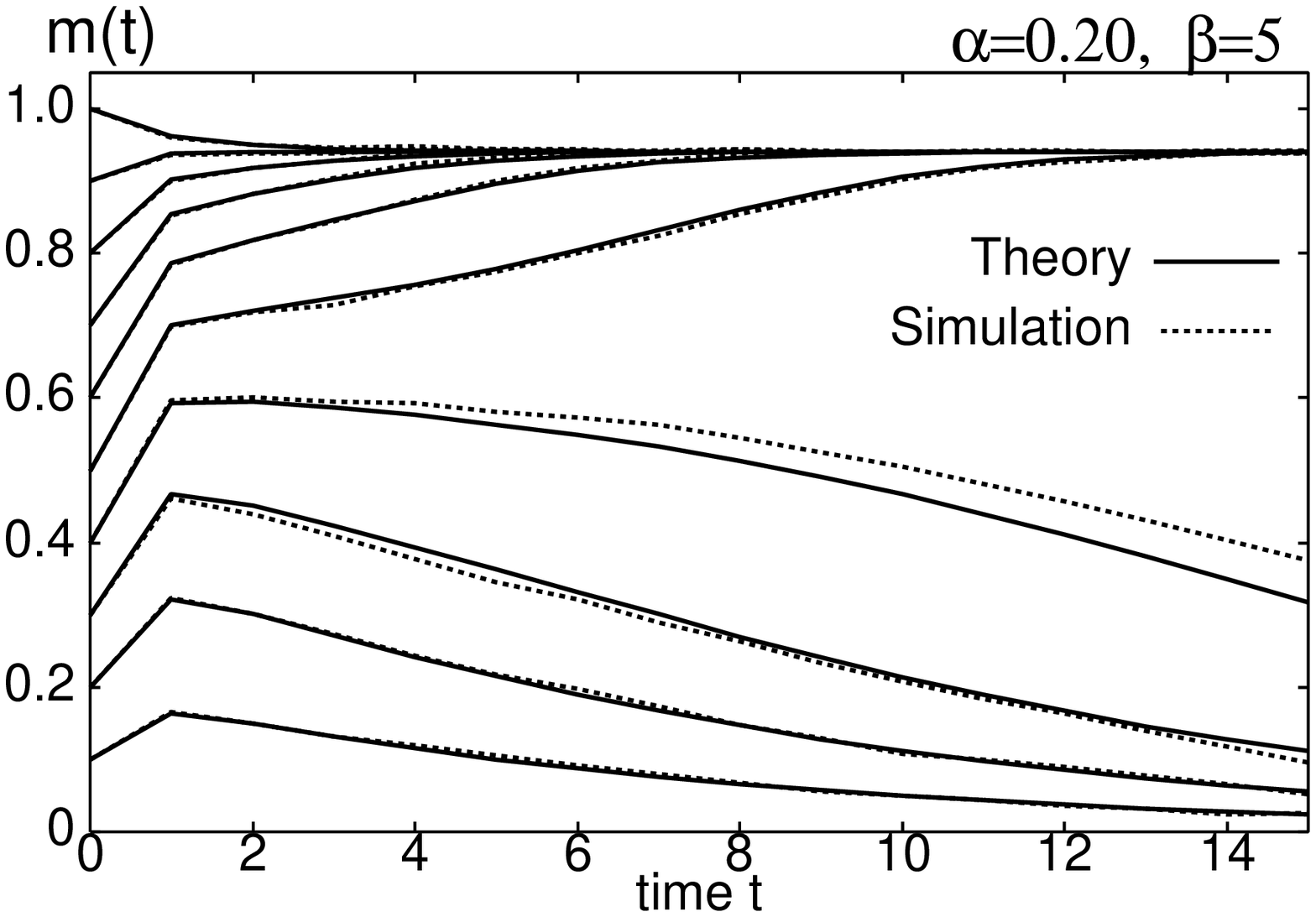} %overlap_0.20_05_100000.eps
  
  \includegraphics[width=130mm]{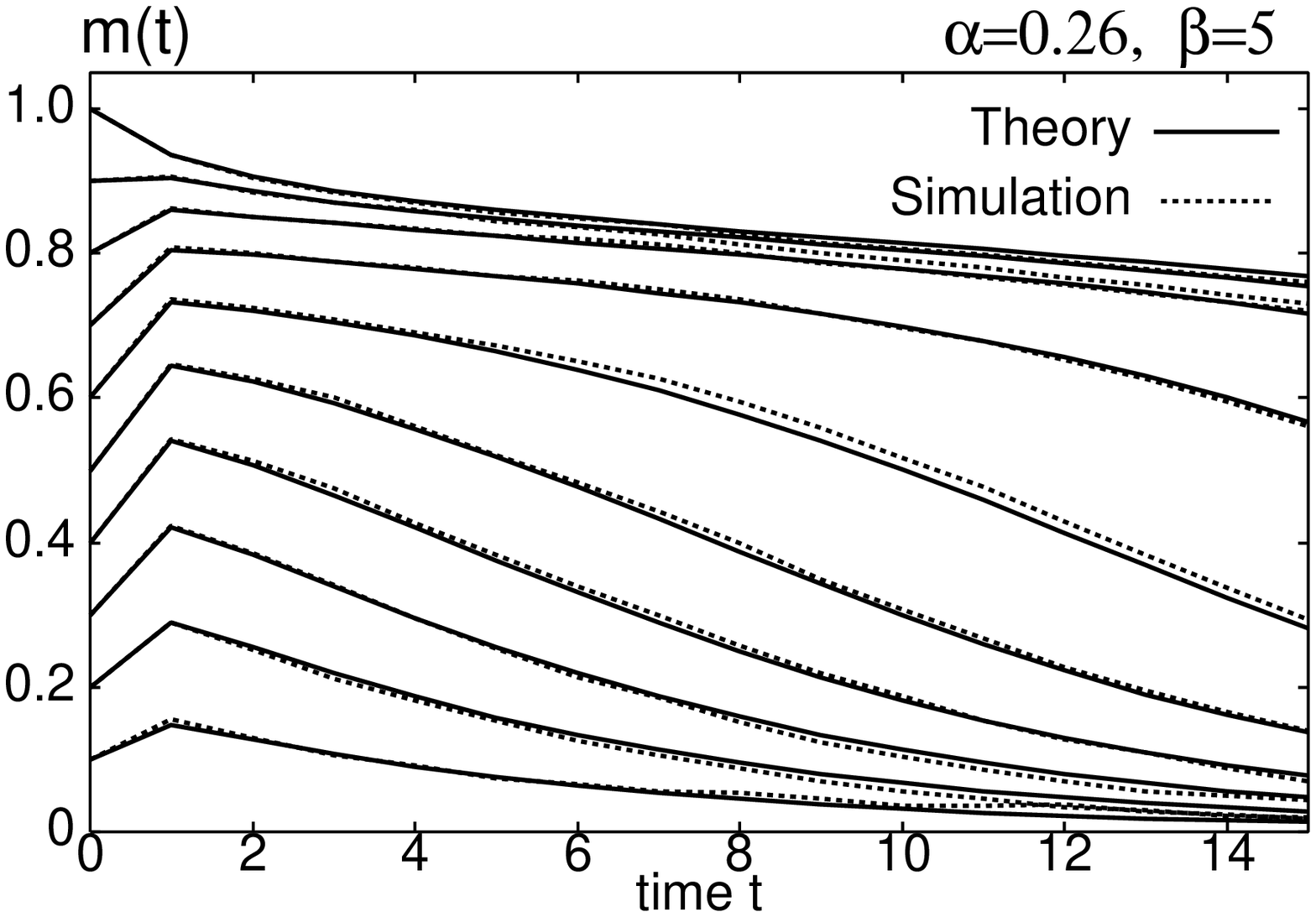} %overlap_0.26_05_100000.eps
 \end{center}
 \caption{Time evolution of overlap $m(t)$ with loading rate
 $\alpha=0.20,0.26$ and inverse temperature $\beta=5$. The solid
 lines denote theoretical results and the broken lines denote results by
 computer simulations with $N=100000$.}  
 \label{fig:overlap}
\end{figure*}

\begin{figure*}
 \begin{center}
  \includegraphics[width=130mm]{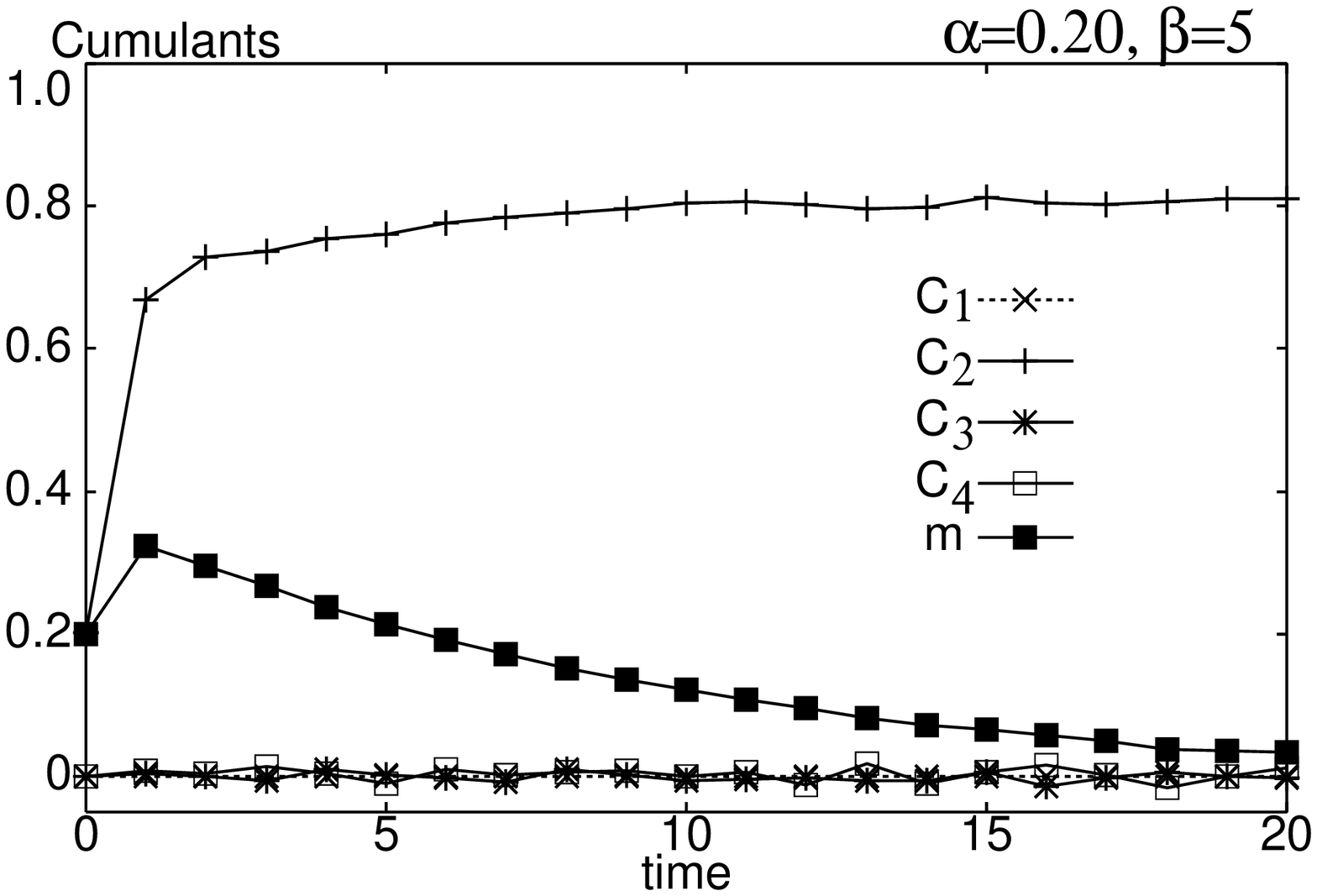} %cumulant_0.20_05_0.20_100000.eps
  \includegraphics[width=130mm]{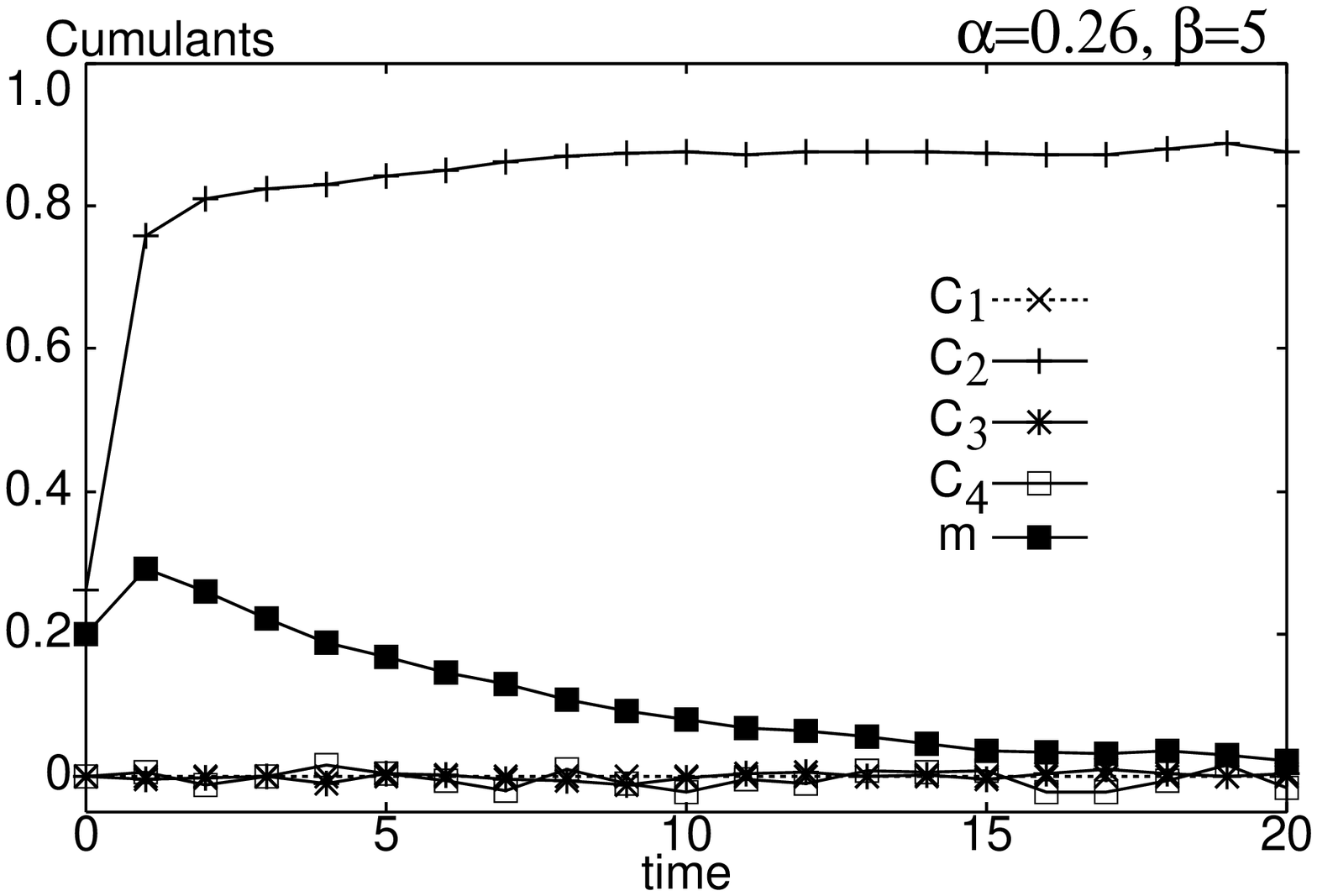} %cumulant_0.26_05_0.20_100000.eps
 \end{center}
 \caption{Time evolution of cumulants $C_1(t),C_2(t),C_3(t),C_4(t)$ and
 overlap $m(t)$ with loading rate $\alpha=0.20,0.26$ and inverse
 temperature $\beta=5$. The initial overlap is $m(0)=0.2$. The solid
 lines denote theoretical results and the broken lines denote results by
 computer simulations with $N=100000$.}  \label{fig:cumulant}
\end{figure*}

\begin{figure}
 \begin{center}
  \includegraphics[width=130mm]{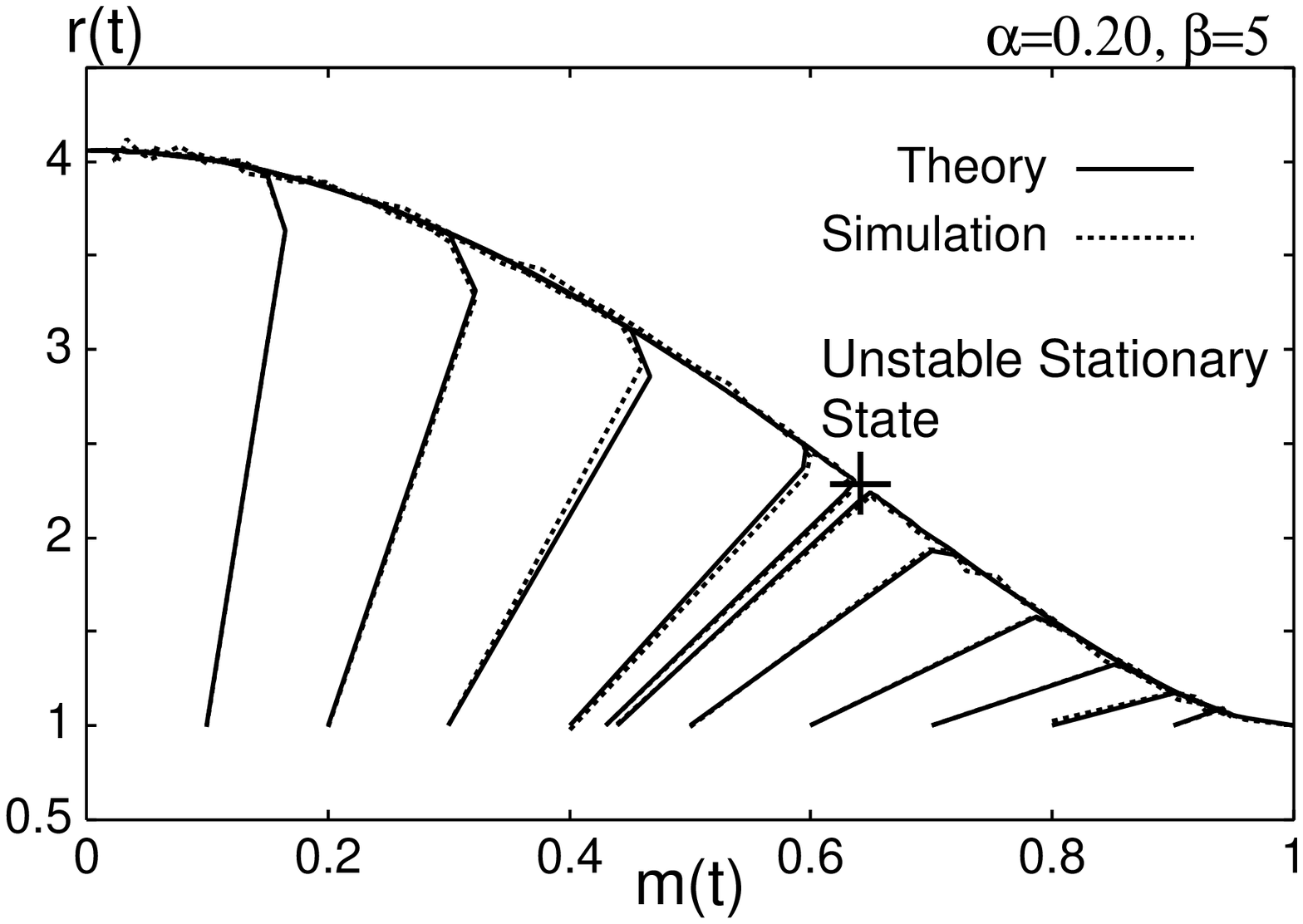} %mr.eps
 \end{center}
 \caption{Overlap $m(t)$ and variance of crosstalk noise $r(t)$ with
 loading rate $\alpha=0.20$ and inverse temperature $\beta=5$. The solid
 lines denote theoretical results and the broken lines denote results by
 computer simulations with $N=100000$. The initial overlap is
 $m(0)=0.10,\cdots,0.40,0.43,0.44,0.50,\cdots,1.0$. } \label{fig:mr}
\end{figure}

\begin{figure}
 \begin{center}
  \includegraphics[width=130mm]{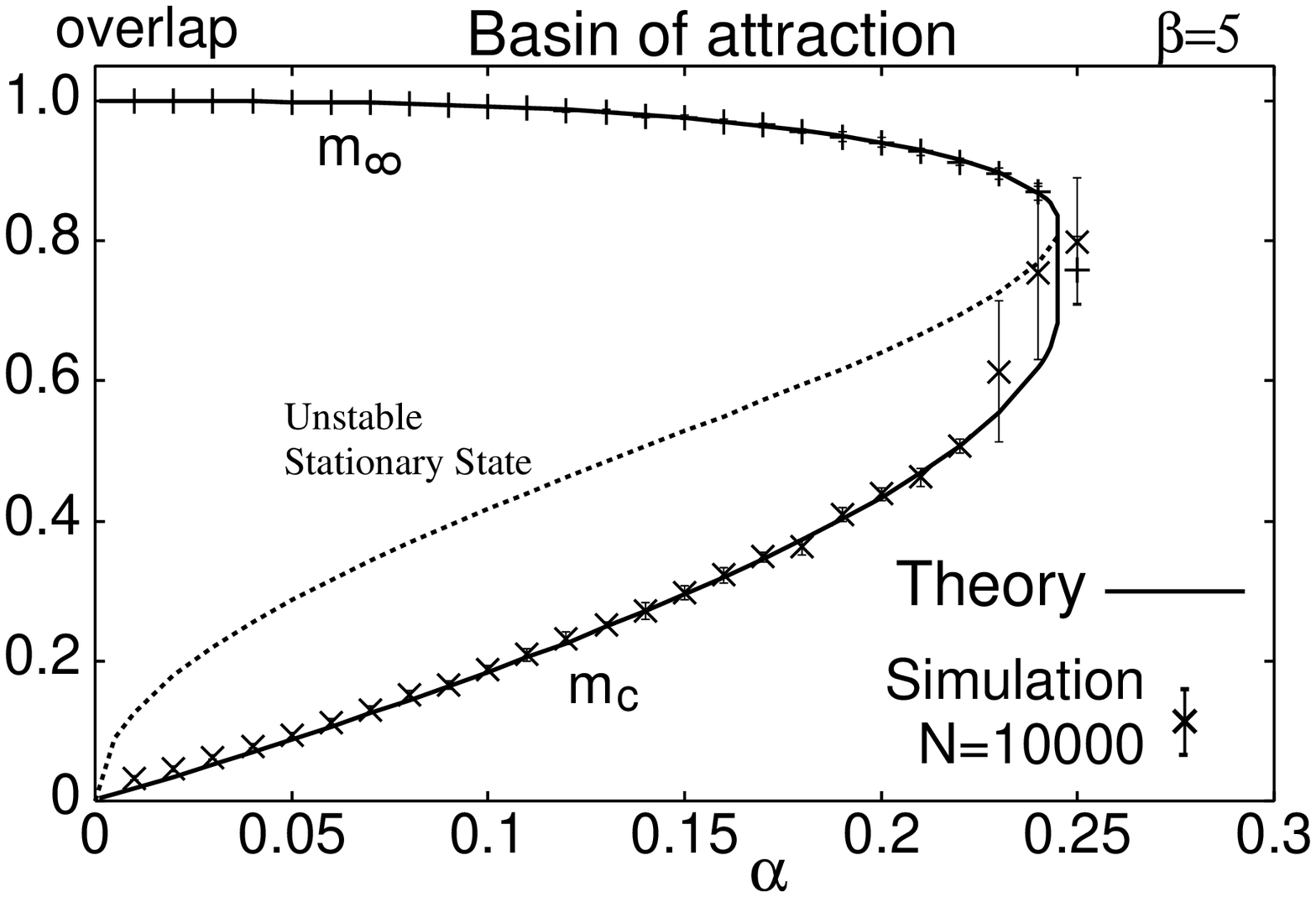} %basin.eps
 \end{center}
 \caption{Basin of attraction (solid line) obtained by theory with
 inverse temperature $\beta=5$.  The error bars denote the median, and
 first and third quartiles over 11 trials by computer simulations with
 $N=10000$.} \label{fig:basin}
\end{figure}

\end{document}